\documentstyle[preprint,aps]{revtex}

\newcommand{\f}{f_q}
\begin{document}
\draft
\preprint{}
\title{Born's principle and Action-Reaction problem}
\author{M. Abolhasani\footnote{E-mail:ablhasan@netware2.ipm.ac.ir}$^{1,2}$, \& M. Golshani$^{1,2}$}
\address{$^1$Institute for Studies in Theoretical Physics and Mathematics
(IPM), P.O.Box 19395-5531, Tehran, Iran.}
\address{$^2$Department of Physics, Sharif University of Technology,
P.O.Box 11365-9161, Tehran, Iran.}
\date{\today}
\maketitle
\begin{abstract}
In a recent paper, A.Valentini tried to obtain Born's principle as
a result of a subquantum heat death, using classical ${\cal H}$-theorem
and the definition of a proper quantum ${\cal H}$-theorem within the framwork of Bohm's
theory. In this paper, we shall show the possibility of solving the problem of
action-reaction asymmetry present in Bohm's theory by modifying Valentini's
procedure. However, we get his main result too.
\end{abstract}
\section{Introduction}
Bohm's theory is a casual interpretation of quantum mechanics that was initially introduced
by de Broglie[1] and then developed by Bohm[2]. This theory is claimed to be
equivalent to the standard quantum theory without having the conceptual problems
of the latter[3]. Yet, there are some difficulties with this theory. One of
these difficulties is the action-reaction (AR) problem[4,5]. In this theory the
velocity of a particle is given by

\begin{eqnarray}
\dot{\vec x}=\frac{\hbar}{m} Im(\frac{\vec\nabla\psi}{\psi})
=\frac{\vec\nabla S}{m}
\end{eqnarray}
where $S$ is the phase of the wave function ($\psi=R e^{i\frac{S}{\hbar}}$). The wave function
$\psi$ itself is a solution of Schr\"odinger equation:

\begin{eqnarray}
i\hbar\frac{\partial\psi}{\partial t}=-\frac{\hbar^2}{2m}\nabla^2 \psi
+V(\vec x)\psi
\end{eqnarray}
By substituting $R e^{i\frac{S}{\hbar}}$ for $\psi$ in (2) , we shall have
the following equations

\begin{eqnarray}
\frac{\partial S}{\partial t}+\frac{(\vec\nabla S)^2}{2m}+V(\vec x)-
\frac{\hbar^2}{2m}\frac{\nabla^2 R}{R}=0
\end{eqnarray}
\begin{eqnarray}
\frac{\partial R^2}{\partial t}+\vec\nabla\cdot(\frac{\vec\nabla S}{m} R^2)=0
\end{eqnarray}
where (3) is the classical Hamilton-Jacobi equation with an additional term, $
Q=-\frac{\hbar^2}{2m}\frac{\nabla^2 R}{R}$, which is called {\it quantum potential}.
By differentiating (1) with respect to $t$ and making use of (3), we obtain

\begin{eqnarray}
m\ddot{\vec x}= -\vec\nabla (Q+V)
\nonumber
\end{eqnarray}
which shows that the $\psi$-wave affects particle's motion through the $-\vec
\nabla Q$ term. To secure the action-reaction symmetry, we expect the presence of a term
corresponding to particle's reaction on the wave, in the wave equation of motion (2).
This term is not present there[5].
\par
Another difficulty with Bohm's theory is Born's statistical principle. In this theory
the field $\psi$ enters as a guiding field for the motion of particles but at
same time it is required by the experimental facts to represent a probability
density through $|\psi|^2$. In this paper we try to obtain Born's statistical principle
and try to solve the AR problem as well.
\par
Our paper is organized as follows: After reviewing Valentini's quantum ${\cal 
H}$-theorem in section 2, we introduce our new quantum ${\cal H}$-theorem in 
section 3 and, finally, we show how the AR problem is solved by our procedure.

\section{quantum $\cal H$-theorem}
Recently, A.Valentini has derived the relation $\rho=|\psi|^2$ as the result
of a statistical subquantum ${\cal H}$-theorem[6]. He looked for a proper 
quantum function $f_N$ that, like the classical N-particle distribution 
function, satisfies Liouville's equation. He considered an ensemble of 
N-body systems which could be described by the wave function $\Psi$ and the 
distribution function $P$. Because $|\Psi|^2$ and $P$ must equalize during 
the assumed heat death, in general $P\neq|\Psi|^2$ and one can write

\begin{eqnarray}
P(X,t)=\f(X,t)\ |\Psi(X,t)|^2
\nonumber
\end{eqnarray}
where $\f$ measures the ratio of $P$ to $|\Psi|^2$ at the point $X$ ($\vec
{x_1}\cdots\vec{x_N}$) at time $t$. Since both $|\Psi|^2$ and $P$ satisfy the
continuity equation ($|\Psi|^2$ due to its being a solution of Schr\"odinger
equation and $P$ by its very definition) one can easily show that $\f$
satisfies the following equation

\begin{eqnarray}
\frac{\partial f_q}{\partial t}+\dot X\cdot\vec\nabla f_q=0
\nonumber
\end{eqnarray}
where $X$ means $\vec{x_1}\cdots\vec{x_N}$ as before. Thus, Valentini
defined his quantum $\cal H$-function in the following way

\begin{eqnarray}
{\cal H}_q=\int\ d^{3N}X\ |\Psi(X,t)|^2\ f_q(X,t)  ln f_q(X,t)
\nonumber
\end{eqnarray}
The only difference with the classical one is that $\f$ is defined in the configuration
space while the classical $f_N$ is defined in the phase space and $d^{3N}X\ d^
{3N}P\rightarrow |\Psi(X,t)|^2\ d^{3N}X$. Valentini used Ehrenfest's coarse-graining
method[7]. One can easily shows that {\large $\frac{d\bar{\cal H}_q}{dt}$}
$\leq 0$, where $\bar{\cal H}_q$ is the coarse-graind ${\cal H}$-function and 
the equality holds in the equilibrium state, where $\bar \f=1$ or $\bar P=\bar
{|\Psi|}^2$. Here $\bar P$ and $\bar{|\Psi|}^2$ are coarse grained forms of 
$P$ and $|\Psi|^2$ respectivily. Valentini termed this process a subquantumic 
heath death. Then, he showed that if a single particle is extracted from the 
large system and prepared in a state with wavefunction $\psi$, its probability 
density $\rho$ will be equal to $|\psi|^2$, provided that $P=|\Psi|^2$ holds 
for the large system. Notice that a one-body system not in quantum equilibrium 
can never relax to quantum equilibrium (when it is left to itself). But any 
one-body system is extracted from a large system.
\par
Here, we want to modify Valentini's procedure so that one can directly obtain Born's
principle for a one-body system. In this case we will be forced to use
Boltzmann's procedure, and that naturally leads to a solution of the AR problem.

\section{An alternative quantum $\cal H$-theorem}
Consider an ensemble of one-body systems. Suppose that all these systems are
in the $\psi$ state and that their distribution function is $\rho$. Furthermore,
suppose that at $t=0$ we have $\rho\neq|\psi|^2$, and define

\begin{eqnarray}
\rho(\vec x,t)=\f(\vec x,t)\ |\psi(\vec x,t)|^2
\end{eqnarray}
\par
In Valentini's procedure the complexity of systems leads to heat death , but our
systems are simple (one-body) ones. Thus, if we want to have heat death, we must
assume that $\f$ satisfies a quantum Boltzmann equation

\begin{eqnarray}
\frac{\partial f_q}{\partial t}+\dot{\vec x}\cdot\vec\nabla f_q=J(f_q)
\end{eqnarray}
where $J(\f)$ is related to the particle reaction on its associated wave. Since $\rho$
satisfies a continuity equation, as the result of its definition, thus (5)
and (6) imply that $|\psi|^2$ does not satisfy the continuity equation any more.
In fact, the continuity equation for $|\psi|^2$ is changed to

\begin{eqnarray}
\frac{\partial |\psi|^2}{\partial t}+\vec\nabla\cdot(\frac{\vec\nabla S}{m}
|\psi|^2)=-\frac{J(\f)}{\f} |\psi|^2
\end{eqnarray}
This means that $\psi$ is a soluation of a nonlinear Schr\"odinger equation. If
we want to have the quantum potential in its regular form, i.e. $(-\frac{
\hbar^2}{2m}\frac{\nabla^2 R}{R})$, we must choose the nonlinear term in a
particular form. The proper selection is

\begin{eqnarray}
i\hbar(\ \frac{\partial}{\partial t}+g(\f)\ )\psi=-\frac{\hbar^2}{2m}\nabla^2 \psi
\end{eqnarray}
where $g(\f)$ is a real function of $\f$. Now with the substitution $\psi=R e^{
i\frac{S}{\hbar}}$ we shall have

\begin{eqnarray}
\frac{\partial S}{\partial t}+\frac{(\vec\nabla S)^2}{2m}-
\frac{\hbar^2}{2m}\frac{\nabla^2 R}{R}=0
\end{eqnarray}
\begin{eqnarray}
\frac{\partial |\psi|^2}{\partial t}+\vec\nabla\cdot(\frac{\vec\nabla S}{m}
|\psi|^2)=- 2\ g(\f)\ |\psi|^2
\end{eqnarray}
By comparing (10) with (7) we have $g(\f)=\frac{1}{2}\ \frac{J(\f)}{\f}$.
Now, we have to select $g(\f)$ in such a way that it leads to the equality of $\rho$
and $|\psi|^2$ (i.e. $\f=1$). Some requirements for such a function is:

1- It must be invariant under $t\rightarrow -t$.

2- It must change its sign for $\f=1$.

If we fine systems for which subquantum
heat death has not occured[8], we shall obtain the actual form
of the $g(\f)$ function. A proper selection is $g(\f)=\alpha(1-\f)$ where
$\alpha$ is a constant. Then, (10) gives (with $\alpha=\frac{1}{2}$)

\begin{eqnarray}
\frac{\partial |\psi|^2}{\partial t}+\vec\nabla\cdot(\frac{\vec\nabla S}{m}
|\psi|^2)=(\f-1)\ |\psi|^2
\end{eqnarray}
Now, consider the right hand side of (11) as the source of $|\psi|^2$ field. At any point of
space where $|\psi|^2< \rho$  (i.e. $\f>$ 1), the source of $|\psi|^2$ is positive
and therefore $|\psi|^2$ increases at that point. On the other hand, at any point
of space where $|\psi|^2 > \rho$  (i.e. $\f<$ 1), the source of $|\psi|^2$ is
negative and therefore $|\psi|^2$ decreases at that point. The variation of $|\psi|^2$
continues until $|\psi|^2$ becomes equal to $\rho$. After that, since both $|\psi|^2$ and
$\rho$ evolve under the same velocity field ($\frac{\vec\nabla S}{m}$), they
remain equal. Here our reasoning is not exact. To prove $\rho\rightarrow|\psi|^2$
exactly, we introduce the following ${\cal H}_q$ function:

\begin{eqnarray}
{\cal H}_q=\int d^3x (\rho-|\psi|^2) ln (\frac{\rho}{|\psi|^2})
\end{eqnarray}
Since $(X-Y)ln(${\Large $\frac{X}{Y}$}$)\geq 0$ for all $X,Y\geq 0$, we have ${\cal
 H}_q\geq 0$ -- the equality being relevant to the case $\rho=|\psi|^2$. If we show that for
the forgoing $J(\f)$ one has {\large $\frac{d{\cal H}_q}{dt}$}$\leq 0$, where again the equality
is to relevant to $\rho=|\psi|^2$ (i.e. $\f=1$) state, we have shown that
$|\psi|^2$ becomes equal to $\rho$ finally. We write (12) in the form

\begin{eqnarray}
{\cal H}_q=\int d^3x |\psi|^2\ (\f-1) ln \f =\int d^3x\ |\psi|^2 G(\f)
\nonumber
\end{eqnarray}
where $G(\f)=(\f-1)\ ln\f $. Now we have for {\large $\frac{d{\cal H}_q}{dt}$}

\begin{eqnarray}
\frac{d{\cal H}_q}{dt}=\int d^3x \left \{\ -\vec\nabla\cdot(\ \dot{\vec x}\ |\psi|^2\ G(\f)\ )
-J(\f)\ \left [\ \frac{G(\f)}{\f}-\frac{\partial G(\f)}{\partial \f}\ \right ]\ |\psi|^2\ \right \}
\end{eqnarray}
where we have done an integration by parts. Passing to the limit of large volumes
and dropping the surface term in (13) leads to

\begin{eqnarray}
\frac{d{\cal H}_q}{dt}=\int d^3x \frac{J(\f)}{\f}\{\f-1+ln\f \}\ |\psi|^2
\nonumber
\end{eqnarray}
The quantity in $\{\}$, is negative for $\f<1$ and positive for $\f>1$. Now,
since the $J(\f)$ (i.e. $\f(1-\f)$) is positive for $\f<1$ and negative for $\f>1$,
the integrand is negative or zero for all values of $\f$ and we have

\begin{eqnarray}
\frac{d{\cal H}_q}{dt}\leq 0
\nonumber
\end{eqnarray}
where the equality holds for $\f=1$. The existance of a quantity that
decreases continuously to its minimum value in $\rho=|\psi|^2$ (i.e. $\f=1$)
guarantees Born's principle.
\section{The action-reaction problem}

In the classical gravity, matter fixes space-time geometry and correspondingly
matter's motion is determined by the space-time geometry. This means that matter and
space-time affect each other. Thus, the action-reaction symmetry is preserved. In fact, the
nonlinearity of Einstein equations is the result of this mutual action-reaction.
In the same way, mutual action-reaction between wave and particle in Bohm's theory,
leads to a nonlinear Schr\"odinger equation. But nonlinearity does not necessarily
mean particle reaction on the wave. Indeed nonlinear terms must contain some
information about particle's position too.
\par
In the last section we showed that Born's principle can be a result of the
presence of a nonlinear term, in the Schr\"odinger equation, with special
characteristics. In this section we want to show that the existence of such
a term solves the AR problem in a satisfactory way.
\par
The nonlinear Schr\"odinger equation with our choice of $g(\f)$ function (i.e. $
\alpha(1-\f)\ $) becomes

\begin{eqnarray}
i\hbar(\frac{\partial}{\partial t}+\alpha(1-\frac{\rho}{|\psi|^2})\ )\psi=
-\frac{\hbar^2}{2m}\nabla^2 \psi
\end{eqnarray}
In this equation, $i\hbar\alpha(1-\frac{\rho}{|\psi|^2})$, indicates particle's
reaction on the wave, and as we expect this nonlinear term contains
information about particle's position through $\rho$. To clarify this point, we write the distribution
function $\rho$ as

\begin{eqnarray}
\rho(\vec x)=\sum_{i=1}^{N}\ \delta(\vec x-\vec x_i)
\nonumber
\end{eqnarray}
where $\vec x_i$ is the position of i-th particle of the ensemble. Then, the
$\rho$-distribution is determined from the position of the chosen particle
and the other members of the ensemble. If we have only one particle instead of an
ensemble of particles, then (14) changes to

\begin{eqnarray}
i\hbar(\frac{\partial}{\partial t}+\alpha(1-\frac{\delta(\vec x-\vec x_i)}
{|\psi|^2})\ )\psi=-\frac{\hbar^2}{2m}\nabla^2 \psi
\nonumber
\end{eqnarray}
Now it is clear that particle's position affects wave evolution, directly.
Thus, there is no AR problem. In an actual experiment we always
confront an ensemble of particles and for such ensembles the
heat death has already occured. Thus the nonlinear term, that is an indicator
of particle's reaction on the wave, is eliminated. This means that the AR problem is a result
of the establishment of Born's principle. In fact this is similar to the
argument that Valentini presents[9] to show that the signal-locality
(i.e. the absence of practical instantaneous signalling) and the uncertainty
principle are valid if and only if $\rho=|\psi|^2$.

\section{conclusion}
We showed that the Born's principle can be the result of the presence of a nonlinear
term in schr\"odinger equation, with special characteristics. Then, we showed
that this nonlinear term is indicator of particle's reaction on the wave. Thus,
in this way, we have not only obtained Born's principle (Bohm's postulate) but
also we have solved the AR problem.
\centerline{\bf Acknowledgments}
The authors would like to thank Prof. J. T. Cushing for useful comments on an 
earlier draft of this paper.

\end{document}